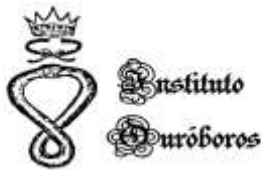

*Colaboran en este número:*

**Bermúdez Vázquez, Manuel**: Dr. en Filosofía. Prof. de Filosofía, Univ. de Córdoba.
**Castillo Arenas, Francisco**: Lcdo. en Historia, DEA. Prof. Geografía e Historia, IES *Mencía López de Haro*, Doña Mencía (Córdoba).
**Enamorado Báez, Santiago Miguel**: Lcdo. en Física. Investigador en prácticas del Centro Nacional de Aceleradores, Sevilla.
**González Gonzalo, Antonio Joaquín**: Dr. en Filosofía y Letras. Prof. de Lengua y Literatura, IES *Felipe Solís*, Cabra (Córdoba).
**Granados Sancho, Mª Araceli**: Lcda. en Filosofía, Univ. de Granada.
**Guerrero Cabrera, Manuel**: Lcdo. en Filología Hispánica. Prof. de Lengua y Literatura, IES *Miguel de Cervantes*, Lucena (Córdoba).
**López Sánchez, Ángel R.**: Dr. en Astrofísica. Contratado Postdoctoral en *Australian Telescope National Facility*, (CSIRO/ATNF), Sydney (Australia).
**Martínez García-Gil, José**: Lcdo. en Biología. Lcdo. en CC. Exactas. Investigador asociado, Programa de Investigación Clínica, CNIO, Madrid.
**Martín-Lorente Rivera, Enrique**: Ingeniero Técnico Industrial, Univ. Córdoba. Empresario.
**Miralles Aranda, Antonio José**: Lcdo. en Biología. Prof. de Ciencias de la Naturaleza, IES *Clara Campoamor*, Lucena (Córdoba).
**Montañez Naz, Sergio**: Dr. en Ciencias, Univ. Autónoma de Madrid. Investigador postdoctoral *Service de Physique Théorique et Mathématique. Université Libre de Bruxelles* (Bélgica).
**Moya Rodriguez, Ana Patricia**: Jefa de la revista *Groenlandia*.
**Muñoz Castillo, Juan Antonio**: Prof. de Historia y Economía, IES *Séneca*, Córdoba.
**Ruiz Gómez, Aarón**: Lcdo. en Física. Contratado Predoctoral del Dpto. de Física Atómica, Molecular y Nuclear, Univ. de Sevilla.
**Ruiz Gómez, Clara Eugenia**: Lcda. Historia. Prof. de Geografía e Historia y responsable de coeducación, IES *Virgen de la Cabeza*. Marmolejo (Jaén)
**Sánchez Fernández, Antonio J.**: Contable, estudiante de Administración y Finanzas, escritor.
**Serrano Castro, Antonio Jesús**: Lcdo. en Derecho y experto en Criminología. Prof. Asociado área de Filosofía del Derecho, Univ. de Córdoba. Abogado.
**Valle Porras, José Manuel**: Prof. de Historia, IES *Ricardo Ortega*, Fuente Alamo (Murcia).
**Vázquez Salas, Carlos**: Prof. de Física y Química, IES *Miguel de Cervantes*, Lucena (Córdoba).
**Ventura Rojas, José Manuel**: Dr. en Historia Univ. de Córdoba. Prof. Asistente de Historia Moderna, Univ. de Concepción (Chile).

# ÍNDICE

**Esta publicación es completamente gratuita y de libre difusión.**

# PARADIGMAS MUERTOS Y PARADIGMAS ASESINOS

Sergio Montañez Naz

**Resumen:** El presente trabajo analiza, mediante determinados ejemplos, la importancia que en el desarrollo de una investigación científica tiene el marco teórico en el que ésta se desarrolla. En particular se presta especial atención a los profundos cambios que ha sufrido a lo largo de la historia el concepto físico de vacío, y se da una explicación pedagógica sobre el importante papel que juega este concepto en la Física Teórica actual.

**Palabras clave:** Filosofía de la Ciencia, Paradigma, Vacío, Principio Antrópico.

Cerramos las fronteras para que no se infiltrase el espíritu de Europa, y Europa se vengó alzando sobre los Pirineos una barrera moral mucho más alta: la muralla del desprecio. Desde fines del siglo XVII, nuestros sabios, nuestros filósofos, nuestros literatos, dejaron casi enteramente de ser leídos y citados. [...] hemos vivido, pues, durante siglos recluidos en nuestra concha, dando vueltas a la noria del aristotelismo y del escolasticismo y desdeñosos (con excepción de pocos paréntesis) del poderoso movimiento crítico y revisionista que impulsó en Europa a las ciencias y a las artes. [...] A causa de esta incompleta conjugación con Europa, nuestros maestros profesaron una ciencia muerta[1].

Con estas palabras, el que a la postre sería premio nobel de Fisiología y Medicina, el navarro-aragonés Santiago Ramón y Cajal, finalizaba, en su discurso de ingreso a la Academia de Ciencias en 1897, una recapitulación de las distintas teorías propuestas desde finales del siglo XVIII sobre las causas del atraso científico español con respecto al resto de Europa. De acuerdo con este punto de vista, el hecho de que nuestro país hubiera hecho tan pocas contribuciones notables a la Ciencia desde el siglo XVII hasta finales del XIX no se debe a que en España hubiera poca actividad científica durante este periodo, sino fundamentalmente a que en las universidades españolas se trabajó dentro del marco teórico de paradigmas científicos obsoletos: en Física predominaba el aristotélico, adoptado por los escolásticos; en medicina, el de Galeno; y en astronomía el de Ptolomeo, por poner algunos ejemplos. Fuera de los marcos teóricos modernos, al margen de las orientaciones filosóficas abiertas en la Europa del siglo XVII y sin practicar el conjunto de métodos de investigación instaurados en el resto del continente, los maestros españoles quedaron al margen de la evolución del pensamiento y desarrollo científico europeo[2]. Su ciencia estaba muerta porque se desarrollaba en el contexto de *paradigmas muertos*.

Esta oscura etapa de la Historia de la Ciencia en nuestro país nos sirve como introducción del breve apunte que pretendemos hacer a continuación acerca de la importancia que, en el desarrollo de toda actividad científica, tiene el marco teórico en el que ésta se desarrolla. De hecho, el punto de vista generalizado en Filosofía de la Ciencia desde los años 60 del pasado siglo es que, para poder acercarnos a una concepción de las ciencias experimentales más realista que la de las corrientes anteriores (positivistas, falsacionistas,...), es fundamental comprender el contexto en el que tiene lugar la actividad y esto incluye tanto el entramado sociológico como el teórico[3]. En este artículo vamos a citar algunos ejemplos en los que la importancia del entramado teórico es tan grande que llega incluso hasta el punto de prevalecer sobre los efectos que otro tipo de influencias podrían provocar.

---

[1] S. RAMÓN Y CAJAL (1897): «Deberes del Estado en relación con la producción científica», Discurso de ingreso a la Real Acadèmia de Ciencias, reimpreso en ID. (1963) *Los tónicos de la voluntad*, Espasa-Calpe, p. 154.
[2] Enrique y Ernesto GARCÍA CAMARERO (1970): *La polémica de la ciencia española*, Alianza Editorial; Enrique GARCÍA CAMARERO (2008): «La Modernización Científica en la España del siglo XIX», Contribución al *Congreso Historia y Ciencia 2008*, Universidad Carlos III de Madrid.
[3] A. CHALMERS (1997): *¿Qué es esa cosa llamada ciencia?*, Siglo XXI.

Antes de comenzar debemos hacer dos puntualizaciones: En primer lugar, hay que señalar que, en mi opinión, este carácter condicionante del marco teórico tiene la capacidad de ir más allá del ámbito puramente científico y de extenderse a todos los aspectos de la vida humana. Así, por ejemplo, estamos acostumbrados a ser espectadores y, a veces, participantes de debates en los que diversos políticos son incapaces de ponerse de acuerdo acerca de las medidas a tomar para resolver o paliar determinado problema o situación y todo ello porque, aunque ellos no sean conscientes, esta situación ni siquiera es la misma en la mente de cada uno de ellos. Recuerdo haber estudiado en el instituto los famosos elementos de la comunicación: el emisor, el receptor, el mensaje, el código, el referente, el canal… En aquella época, desde mi ingenuidad, pensaba que uno de estos elementos era mucho más importante que los demás: se trataba del código. Si dos interlocutores utilizan el mismo código, pensaba yo, tienen necesariamente que entenderse. Sin embargo, hoy puedo decir que la inmensa mayoría de los malentendidos que he presenciado en mi vida, de las dificultades que he tenido como estudiante (y que tengo actualmente como investigador) para entender algún texto o discurso y de las dificultades que he tenido como profesor para enseñar Ciencia a los alumnos/as, se deben fundamentalmente a que emisor y receptor *están considerando referentes distintos*.

En segundo lugar, quisiera señalar que, a pesar de esta casi-universalidad en la que creo, he decidido ajustarme en este análisis sólo a algunos aspectos del quehacer científico y, en concreto, de las ciencias naturales, fundamentalmente porque considero que es un ámbito muchísimo más simple de estudio; a pesar de que ya es, a su vez, complejísimo y vastísimo.

Es habitual encontrarse en libros de texto de Ciencia, fundamentalmente de bachillerato, capítulos dedicados a la naturaleza de las ciencias naturales, en los que se afirma que fue a partir del momento en que disciplinas como la Física, la Química y la Biología iniciaron su camino de ciencia experimental cuando se consolidaron como «auténticas ciencias», abandonando el «oscurantismo» y el «esoterismo» que las rodeaban en épocas anteriores. De alguna manera, esta imagen simplificada está presuponiendo implícitamente que los hechos en los que descansan las ciencias experimentales son anteriores a la teoría e independientes de ella, de tal manera que *bastaría simplemente un estudio experimental serio y cuidadoso para obtener información que constituya un fundamento firme y confiable del conocimiento científico*. Sin embargo, tras un simple estudio introductorio un poco más serio de la Historia de la Ciencia, uno se da cuenta de que la realidad es mucho más compleja. El motivo del atraso científico español no tiene nada que ver con que los científicos españoles en los siglos XVII, XVIII y XIX fueran o no lo suficientemente cuidadosos y serios en sus experimentos[4].

Un contraejemplo bastante impactante que ilustra la falsedad de la proposición en cursivas del anterior párrafo lo encontramos en el desarrollo de la química del siglo XIX. En el contexto de la teoría atómica de la combinación química, era muy interesante estudiar la validez de la famosa hipótesis de William Prout, que afirmaba que el átomo de hidrógeno representaba el ladrillo con el que estaban construidos los otros átomos[5]. A la luz de los conocimientos actuales Prout no iba muy descaminado, puesto que el átomo de hidrógeno posee en su núcleo solamente un protón, mientras que los átomos del resto de los elementos tienen varios protones: el helio dos, el litio tres, etc… Durante décadas, algunos de los químicos más prestigiosos hicieron un trabajo experimental formidable para medir con una gran precisión las masas atómicas de los distintos elementos y comprobar así si eran múltiplos de la del átomo de hidrógeno. Lo que no sabían estos investigadores es que existen varios isótopos de cada elemento, cada uno de ellos con el mismo número de protones en el núcleo (número que caracteriza al elemento), pero con distinto número de neutrones y, por tanto, con distinta masa atómica. Lo que habían medido estos químicos con tanta precisión no eran masas exactas de átomos individuales, sino el promedio de la masa de los distintos isótopos de las muestran con las que trabajaban[6]. Este promedio depende tanto de las masas de cada isótopo como de la proporción de cada uno de ellos que podemos encontrar en la naturaleza; y este último dato puede considerarse contingente en tanto que depende, más que de las leyes fundamentales de la naturaleza, de las características específicas de formación y edad del planeta Tierra, de nuestro sistema solar y de nuestra galaxia[7].

---

[4] Un ejemplo clásico lo podemos encontrar en los trabajos de A. KOYRE (1981): *Estudios galileanos*, Siglo XXI; y T. KUHN (1971): *La estructura de las revoluciones científicas*, FCE. Para los aristotélicos, que creían que un cuerpo pesado se desplazaba, por su propia naturaleza, de una posición superior a una más baja hasta llegar a un estado de reposo natural, un cuerpo que se balanceaba simplemente estaba cayendo con dificultad. Galileo, por otra parte, al observar el cuerpo que se balanceaba, vio un péndulo, un cuerpo que casi lograba repetir el mismo movimiento, una y otra vez. Gracias a este nuevo punto de vista Galileo observó también otras propiedades del péndulo y construyó muchas de las partes más importantes y revolucionarias de su nueva mecánica. Por tanto, es a Galileo a quien hay que atribuir el mérito de conseguir este original cambio de visión. Pero nótese que este mérito no se manifiesta en este caso como observación más exacta u objetiva del cuerpo que se balancea. En cuanto a la capacidad descriptiva se puede decir que la percepción aristotélica tenía la misma exactitud.

Este último ejemplo nos sirve para introducir el segundo concepto sobre el que gira este artículo: el de *paradigma asesino*. Un cambio de paradigma puede bastar para echar por tierra todo un trabajo experimental extenso y rigurosísimo, trabajo experimental que está bien hecho, pero que *ha dejado de ser relevante*. A continuación vamos a hablar de un concepto físico, el de «vacío» (que ha sufrido profundas transformaciones a lo largo de la Historia de la Ciencia) y de cómo estas transformaciones han ido cambiando lo que los científicos consideran relevante o no. El motivo por el que he elegido este ejemplo es porque este concepto ocupa un papel central en la Física Teórica actual y está relacionado con un cambio reciente de paradigma que está resultando ser bastante controvertido.

El concepto de vacío ha sido objeto de múltiples discusiones filosóficas desde la Grecia antigua. La hipótesis, sostenida por Aristóteles, entre otros, de que es imposible crear el vacío fue sometida a diversos experimentos tanto en la Antigüedad como en el mundo islámico medieval. La creación de vacío resultaba una empresa harto difícil, dado que se observó que el aire tendía a ocupar siempre el máximo espacio posible y que cuanto más aire se extraía de un recipiente más difícil resultaba seguir extrayendo. De hecho, el primer estudio empírico del que se tiene constancia en el que se consiguió producir el vacío data de mediados del siglo XVII, cuando italiano Gasparo Berti inventó el barómetro de agua. Para entender el fundamento físico de esta «producción de vacío» nos vamos a valer del barómetro de mercurio, inventado unos años más tarde por el también italiano Evangelista Torricelli. Éste consta de una cubeta llena de mercurio y de un tubo (pongamos que de unos 80 cm) cerrado en uno de sus extremos (llamémosle E1, y al extremo abierto E2). Si se llena también el tubo de mercurio, se tapa el extremo abierto del mismo, se le sumerge en la cubeta y se coloca en posición vertical con el extremo E1 en la parte superior, el dispositivo queda como se muestra en la figura 1. Se observa que, al destapar el extremo E2, el mercurio del tubo desciende unos centímetros, dejando en la parte superior un espacio vacío[8]. La ecuación fundamental de la hidrostática nos dice que los puntos A y B se encuentran, por estar a la misma altura, a la misma presión. Por tanto, el peso de la columna de mercurio que hay sobre el punto A da lugar a que se ejerza una presión en A igual a la presión atmosférica. Supongamos que ahora la presión atmosférica disminuyera ligeramente[9]: entonces claramente la columna de mercurio sobre A va a ser menos alta, aumentando de tamaño el espacio vacío entre la superficie de la columna y el final del tubo.

---

[5] W. PROUT (1815): «On the relation between the specific gravities of bodies in their gaseous state and the weights of their atoms», *Annals of Philosophy*, 6, pp. 321-330; ID. (1816): «Correction of a mistake in the essay on the relation between the specific gravities of bodies in their gaseous state and the weights of their atoms», *Annals of Philosophy*, 7, pp. 111-13.
[6] I. LAKATOS (1970): «Falsification and the Methodology of Scientific Research Programmes», en ID y A. MUSGRAVE (eds.): *Criticism and the Growth of Knowledge*, Cambridge University Press, pp. 91-195.
[7] Este programa de medir las masas atómicas es uno de los ejemplos de Lakatos de «programa de investigación degenerado». Más tarde se propuso separar los distintos isótopos mediante métodos basados en la centrifugación, de tal manera que este ejemplo también le ha servido a Lakatos para ilustrar la idea de que no siempre es una opción correcta abandonar un programa de investigación que esté degenerando.

Este comportamiento de los fluidos que parece tan sorprendente resulta bastante natural si nos situamos en el marco de la teoría cinético-molecular. Según esta teoría, los gases están formados por moléculas que se mueven en un espacio vacío, cada una con distintas velocidades (con una velocidad promedio que depende de lo caliente que esté el gas) y que chocan unas con otras y con las paredes del recipiente en el que está contenido. La presión no es más que la fuerza por unidad de superficie que se ejerce como consecuencia de estos choques. Por eso la única posición de equilibrio que puede tener una pared que separa dos gases (véase figura 2) es aquella para la cual los dos gases tienen la misma presión. Si el lado derecho del recipiente estuviera vacio, se ejercería una fuerza muy grande sobre la pared que la obligaría a moverse en el sentido de hacer desaparecer el vacío. Si no hubiera pared, nada impediría que aquellas moléculas de aire que tengan una componente horizontal de la velocidad distinta de cero invadieran la parte derecha del recipiente, expandiéndose el gas y desapareciendo el vacío.

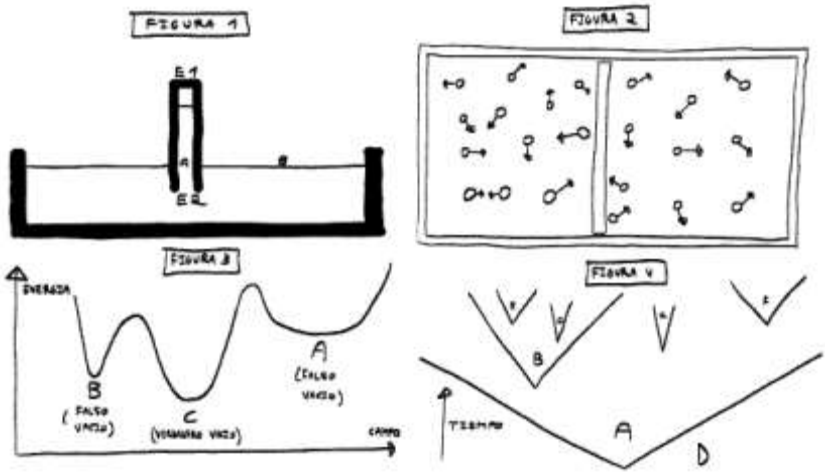

Por tanto, el vacío, aunque experimentalmente resulta muy difícil de realizar[10], desde el punto de vista conceptual no acarrea ningún problema: el vacío es simplemente la ausencia de partículas. En otras palabras: *en este paradigma el vacío no constituye problema físico alguno*. Trivialmente, si no hay partículas no hay sistema físico que estudiar.

Pero las cosas no son tan sencillas. La teoría de la relatividad nos dice que no existen las interacciones instantáneas entre partículas. Para que lo que ocurre en un lugar pueda afectar a las partículas que hay situadas en otro lugar es necesario que pase un tiempo, como mínimo, igual al que tardaría la luz en llegar a ellas. Este simple postulado acarrea un cambio radical a la hora representarnos los sistemas físicos: implica que conceptos como el de campo (campo electromagnético, campo gravitatorio) pasan, de ser meros artificios matemáticos para calcular los detalles de la interacción entre partículas, a tener entidad física propia. La aplicación de la teoría cuántica a estas nuevas entidades físicas ha dado lugar al paradigma de la teoría cuántica de campos (que sigue vigente hasta la actualidad). Según esta teoría, los elementos fundamentales de los que está hecha la materia son una serie de campos que lo impregnan todo. En este paradigma las partículas no son más que excitaciones (estados de vibración) de los campos que se propagan de un lugar a otro. Así, por ejemplo, los fotones son las excitaciones del campo electromagnético y los electrones son las excitaciones de un campo «electrónico». Si tenemos en cuenta la mecánica cuántica, estos campos sólo pueden excitarse en niveles energéticos discretizados, de la misma forma que los niveles de energía que tienen los átomos son discretos. Este es el motivo por el que puede existir un electrón, dos electrones, etc… pero nunca un electrón y medio. ¿Qué sería el vacío entonces? Pues la ausencia de partículas, que en este caso significa que los campos se encuentran sin excitaciones (en su estado de menor vibración)[11]. Este estado puede ser un verdadero vacío (un verdadero estado de menor energía), o un falso vacío. En este último caso el campo posee una mínima vibración en torno a un valor cuya energía asociada es menor que la de sus valores vecinos, pero que no es un mínimo absoluto (véase figura 3).

---

[8] La altura de la columna de mercurio en el tubo, medida desde la superficie del mercurio de la cubeta, es, en condiciones normales y al nivel del mar, de 76,0 cm.

[9] Nótese que no hace falta esperar a una borrasca. El cuñado de Torricelli, Blas Pascal, consiguió esto llevando el barómetro a la cima de una montaña de 1.000 metros de altura.

[10] Ni siquiera el vacío que se produce en el barómetro de Torricelli es un verdadero vacío puesto que en él hay una pequeña cantidad de átomos de mercurio moviéndose, radiación electromagnética, etc.... No obstante, para determinados experimentos podemos considerar que es una buena aproximación de vacío.

Con este nuevo paradigma, *el vacío se convierte en un problema físico formidable*: no es nada trivial determinar cuáles son los posibles estados de mínima vibración en los que pueden encontrarse los campos. Y, además de formidable, es de máximo interés; puesto que, como podemos observar en la figura 3, las partículas van a tener características muy distintas según en torno a cuál de los posibles vacíos se encuentra el campo. Por ejemplo, en el vacio B cuesta más que el campo se excite (porque el valle es más estrecho) con lo que la masa de las partículas va a ser mayor.

La cosa se complica más todavía si tenemos en cuenta que, de acuerdo con la relatividad general, el mismo espacio-tiempo es, por sí mismo, también un campo (el campo gravitatorio). El conjunto de todos estos campos posee una densidad de energía de vacío distinta de cero que es la responsable de que el universo se encuentre actualmente en expansión acelerada. Esta energía aparece en las ecuaciones de evolución cosmológica en forma de una constante, denominada constante cosmológica. Tratar de entender por qué esta constante toma el valor observado supone uno de los grandes problemas sin resolver de la física teórica actual.

Hasta hace unos años, el punto de vista generalizado entre los físicos teóricos era que el problema del vacío se iba a poder resolver en el contexto de una «teoría del todo»: una teoría que explique la totalidad de la materia y de las interacciones que observamos en la naturaleza sin necesidad de ajustar ningún parámetro. De hecho, el marco dominante en el que están trabajando los físicos teóricos, la teoría de cuerdas, parecía cumplir estas características: ¡una teoría del todo que no posee parámetros libres! Pero ocurre que actualmente no entendemos bien la estructura de vacios de esta teoría. Por lo que sabemos hasta ahora, resulta que esta teoría aparentemente tiene un gran número de vacíos metaestables (quizás $10^{1000}$, ¡o incluso más!) y cada uno de ellos corresponde a un universo posible, con distintas colecciones de partículas e interacciones entre ellas, y con distinta constante cosmológica. Al conjunto de todos ellos se le denomina el *string theory landscape*. Muchos de estos vacíos del *landscape* corresponden también a un universo en expansión, de manera que, para cuando comience en alguna pequeña región de la parte del universo situada en el vacio A el decaimiento a otro vacio B de menor energía, ¡la región del universo correspondiente a A ya es suficientemente grande como para poder ser todo el universo que observamos! (véase figura 4).

---

[11] El principio de indeterminación impide que pueda haber un estado sin vibración alguna.

Muchos físicos teóricos, entre los que podemos destacar a Leonard Susskind y Andrei Linde[12], piensan que esta característica de la teoría de cuerdas es una ventaja, ya que permite una explicación antrópica tanto del valor observado de la constante cosmológica como de las partículas y sus interacciones: *en el universo puede haber muchísimas regiones correspondientes a los distintos vacíos posibles (distinta constante cosmológica, distintos tipos de partículas que interaccionan según leyes físicas distintas, etc..), pero nosotros nos encontramos en aquella región en la que las leyes de la física han permitido la vida inteligente.*

Este punto de vista ha tenido tan buena acogida en algunos sectores de la comunidad de teóricos de cuerdas que inmediatamente ha suscitado la reacción de otros (entre los que destaca el premio Nobel de Física de 2004, David Gross) por considerar esta idea del *landscape* antrópico demasiado peligrosa: se trata de una idea difícil (posiblemente imposible) de testear. Existe la posibilidad de que, sea cual sea el resultado de los distintos experimentos sobre el comportamiento de la naturaleza, la teoría de cuerdas siempre tenga un vacío metaestable (con tiempo de decaimiento del orden de la edad del universo) que explique el comportamiento observado. Si esto es así no se trataría de una teoría científica, en el sentido popperiano de que *no sería falsable*.

El peligro del paradigma del *landscape* antrópico se debe a que, en cierto sentido, es un paradigma asesino, como el de los isótopos del ejemplo anterior. Sugiere que los valores observados de determinadas constantes físicas, como la constante cosmológica, ¡podrían tener tanta relación con las leyes fundamentales de la naturaleza como las masas atómicas promedio que midieron los químicos en el siglo XIX!, con el consiguiente peligro de que no se busque explicación fundamental a estos valores.

Comenzamos este artículo citando unas palabras del premio nobel de Medicina y Fisiología de 1906 y lo vamos a finalizar con las de otro premio Nobel, el de Física de 2004, acerca de la utilidad en Ciencia de evitar los argumentos antrópicos:

> The main reason I think people take this anthropic argument seriously is the value of the cosmological constant: how do we explain that? [...] It is a small number [...] but we have explained such a small numbers before. Remember Dirac in 1937 [...] was worried about why is the Planck mass so much bigger than the proton mass ($M_{proton}/M_{Planck}=10^{-19}$) [...] He did not invoke anthropic arguments by saying that if this number was of order 1 we would not be here! Instead he used it to make a prediction: he suggested that the ratio $M_{proton}/M_{Planck}$ was related to the size of the universe in atomic units [...] 30 years ago the Dirac's $M_{proton}/M_{Planck}$ tiny ratio *was explained* by QCD's log running of the strong interaction coupling from the unification scale[13].

Si Gross tiene razón con sus advertencias o no, sólo el tiempo nos lo dirá[14].

---

[12] L. SUSSKIND (2003): «The Anthropic landscape of string theory», en B. CARR (ed.): *Universe or multiverse?* 247-266, e-Print: hep-th/0302219; A. LINDE (2007): «Eternal Inflation and String Theory Landscape», en *Strings 07 Madrid* <http://www.ift.uam.es/strings07/040_scientific07_contents/videos/linde.mp4>.

[13] D. GROSS (2007): «Perspectives», en *Strings 07 Madrid*
<http://www.ift.uam.es/strings07/040_scientific07_contents/videos/gross.mp4>.

[14] **Agradecimientos**: Quisiera dar las gracias a Silvia Cid, Guillermo Ballesteros y Carlos Hoyos por sus comentarios y sugerencias, que han sido de gran utilidad para la mejora del texto. También quisiera agradecer a Enrique García Camarero el haberme dado la oportunidad de acercarme a su trabajo, de cuya lectura me surgió la idea de escribir este artículo, y diversas tertulias de las que he aprendido mucho. The work of SM is supported by the Belgian Fonds de la Recherche Scientifique (FNRS).